# Interference Coordination Schemes for Wireless Mobile Advanced Systems: A Survey

Maissa Boujelben[1], Sonia Benrejeb[2], Sami Tabbane[3]
[1]*(MEDIATRON, Sup'Com, Tunisia)*
[2]*(MEDIATRON, Sup'Com, Tunisia)*
[3]*(MEDIATRON, Sup'Com, Tunisia)*

***Abstract:*** *Mobile communication networks have witnessed a perpetual evolution since their launching as voice only networks. In the few past years, the focus was addressed to high data rate networks that offer high quality of service. The lately released LTE-Advanced network was the first to completely fulfill 4G requirements. However, performance gains remain limited due to severe interference levels. This paper is a survey upon the evolution of interference mitigation solutions from Release 8 to Release 11. This problem was addressed since earlier releases by coordinating transmission/reception among different cells. Many enhancements were then carried out by each subsequent specification. The originality of this work resides on the comparison between the different schemes and the perspective of developing a new interference coordination method available for 4G and beyond systems with small cells deployment. Several similarities are noted and differences on performance impact are highlighted.*

## I. INTRODUCTION

Third Generation Partnership Project (3GPP) work on the Evolution of the 3G Mobile System started with the Radio Access Network (RAN) Evolution Work Shop in November 2004 in Toronto, Canada. The Work Shop was open to all interested organizations who presented many contributions with views and proposals on the evolution of the Universal Terrestrial Radio Access Network (UTRAN). With the conclusions of this Work Shop and with broad support from 3GPP members, a feasibility study on the Universal Terrestrial Radio Access (UTRA) and Universal Terrestrial Radio Access Network (UTRAN) Long Term Evolution (LTE) was started in December 2004. The objective was "to develop a framework for the evolution of the 3GPP radio-access technology towards a high-data-rate, low-latency and packet-optimized radio-access technology". The study focused on supporting services provided from the PS domain. [1]

Release 8 was frozen in December 2008 and this has been the basis for the first wave of LTE equipment. Long Term Evolution (LTE) and its Evolved Packet switched System (EPS) architecture, commonly known as System Architecture Evolution (SAE), were defined and included in 3GPP Release 8 (see Fig. 1). LTE specifications are very stable, with the added benefit of enhancements having been introduced in all subsequent 3GPP Releases.

SAE offers many advantages over previous and current 3G systems, e.g. all-IP architecture supporting both IPv4 and IPv6, reduced latency, adaptability, scalability, robust radio access technology, reduced deployment and maintenance costs, etc. However, LTE as standardized in 3GPP Release 8 is not a full 4G standard despite it offers significant performances improvement. The driving force to further develop LTE towards LTE–Advanced - LTE Release10 was to increase capacity, provide higher bitrates in a cost efficient way and, at the same time, completely fulfill the requirements set by The International Telecommunication Union—Radio Communication Sector (ITU-R) for International Mobile Telecommunications- Advanced (IMT Advanced), also referred to as 4G. IMT-Advanced offers high quality of services and supports enhanced peak data rates in order of 100 Mbit/s for high mobility (up to 350 km/h) and 1Gbit/s for low-mobility environments (up to 10 km/h).

It is worth highlighting that LTE-Advanced was designed to be backward compatible with existing cellular architecture for both operator and user interest. On the one hand, it is very important for an operator to softly integrate a new system without dropping previous investments. On the other hand, this backward compatibility allows users to benefit from already acquired terminals.

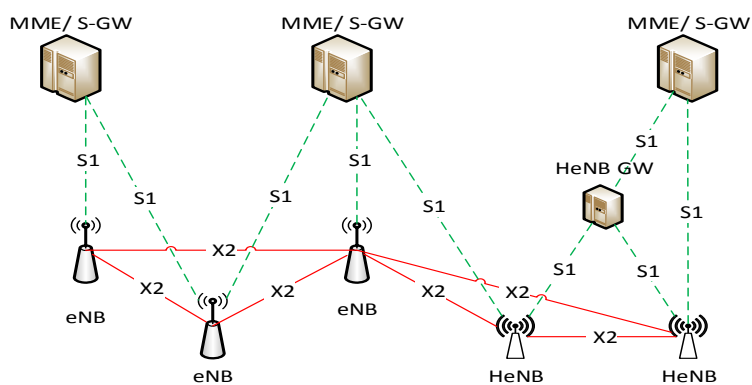

Figure. 1. Overall SAE architecture

In LTE systems, the possibility for efficient heterogeneous network planning – i.e. a mix of large and small cells - is increased by introduction of Low Power Nodes (LPN). At the end of 2008, Femto Forum joined 3GPP as a Market Representation Partner (MRP) to work to support residential and enterprise deployments with open or closed access called Home eNodeB (HeNB). The main purpose behind supporting multi-layer deployment is the traffic offloading from Macro cells or eNB. The E-UTRAN architecture may deploy a HeNB Gateway (HeNB GW) to allow the S1 interface between the HeNB and the EPC to scale to support a large number of HeNBs (see Fig. 1).

The main purpose of designing LTE and LTE-Advanced systems was increasing data bit rate and enhancing the spectral efficiency especially at cell edges. However, this evolution used to be braked by severe inter cell interference. This problem is essentially caused by two sources: the first is the frequency reuse factor equal to 1 in LTE systems. This means that all neighbor eNodeBs (eNB) use the same frequency channels which will generate more interference. The second issue is the heterogeneous deployment of LTE networks comprising of conventional Macro cell base stations overlaid with LPNs. The often random and unplanned location of these access points can cause severe interference problems especially for cell edge users.

This problem was typically addressed by coordinating base-station transmissions to minimize interference. Towards this direction, 3GPP LTE standard has introduced Inter Cell Interference Coordination (ICIC) methods since Release 8 specifications. ICIC was developed to deal with interference issues at cell-edge and mitigates in-terference on traffic channels only. These limitations were bypassed with Release 10 specifications which intro-duced enhanced Inter Cell Interference Coordination (e-ICIC). Enhancements were brought to deal with interference issues in Heterogeneous Networks and mitigate interference on traffic and control channels.

Coordination techniques developed in Release 8, 9 and 10 are semi-static. This makes the achievable benefit limited as it is not able to follow the dynamic interference condition caused by dynamic scheduling. Dynamically coordinating the transmission and reception of signals at multiple cells could lead to further performance improvement. Coordinated Multi Point transmission and reception (COMP) was brought by 3GPP LTE-Advanced Release 11 and its performance benefits were well studied and proven.

In this work, we focus on studying interference mitigation techniques as specified in 3GPP LTE and LTE-Advanced standards, and we address the performance gain comparison of the different schemes. Therefore, we first give an overview of inter cell interference mitigation standards. Then, we compare the different schemes and their performance impact. Finally, we give conclusions.

## II. INTERFERENCE COORDINATION SCHEMES

As mentioned before, LTE standard is designed for frequency reuse 1 (to maximize spectrum efficiency), which means that all the neighbor cells are using same frequency channels and therefore there is no cell-planning to deal with the interference issues. Thus, there is a high probability that a resource block scheduled to cell edge user, is also being transmitted by neighbor cell, resulting in high interference and eventually low throughput or call drops. Besides, heterogeneous networks require some sort of interference mitigation, since small cells and macro cells are overlapping in many scenarios.

Inter cell interference mitigation schemes can be grouped under two categories: static and dynamic schemes (see Fig. 2).

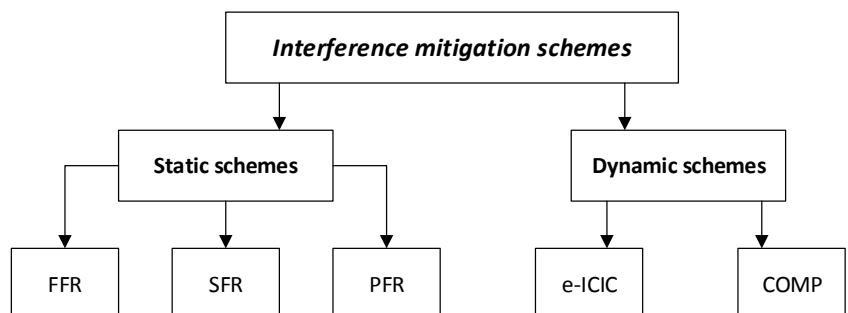

Figure. 2. Interference mitigation schemes

### 2.1. STATIC SCHEMES: RELEASE 8/9 ICIC

Inter-cell interference coordination was first introduced by 3GPP Release 8 LTE to deal with interference is-sues at cell-edge by mitigating interference on traffic channels only. It has the task to manage frequency radio resources (notably the radio resource blocks) such that inter-cell interference is kept under control. ICIC is inherently a multi-cell Radio Resource Management (RRM) function that needs to take into account information (e.g. the resource usage status and traffic load situation) from multiple cells. The preferred ICIC method may be different in the uplink and downlink. ICIC function is located in the eNB [2].

The coordination between cell sites is achieved by exchanging messages between eNBs over X2 inter-face. Frequency-domain ICIC over downlink in Rel-8 is based on controlling the downlink cell power for resources. This is achieved by sending Relative Narrowband Transmit Power (RNTP) message as often as every 200 ms. This message contains information whether or not the frequency time resource is limited by transmit power. When a neighbor eNB listens to this message, it can avoid scheduling on the indicated resources.

Two messages are defined for uplink interference coordination: High Interference Indicator (HII) and over-load indication (OI) exchanged between eNBs as often as every 20 ms.

HII is used to communicate, on which frequency time resources and eNB is going to schedule cell edge users. By listening to this message, a neighbor eNB can avoid scheduling cell edge users in the indicated resources. This can, therefore, result in reduced uplink interference for both of the cells. The action to be taken by an eNB when it receives HII message is implementation specific.

An eNB sends Overload Indicator message to indicate the level of interference experience in different frequency time resources to neighbor eNB. Three levels of interference are defined: Low, Mid and High. When an eNB receives overload indicator message, it can change the scheduling pattern to free the resources indicated in the overload indicator message, therefore, reducing the interference for cell edge users.

Static schemes usually fall into one of three broad categories: traditional hard fractional frequency reuse (FFR), soft frequency reuse (SFR) [3], and partial frequency reuse (PFR) [4].

### 2.1.1. FRACTIONAL FREQUENCY REUSE

To avoid the limitations of the traditional frequency reuse schemes, the fractional frequency reuse scheme is introduced to obtain a frequency reuse factor between 1 and 3. FFR divides the whole available frequency bands into two groups, one for cell center users and the other for cell edge users. In the first group, resources are used with a reuse factor equal to one. This means that all cell center users in adjacent cells can be scheduled with the same resources. The second group is contrarily divided into three subsets, which allows a reuse factor equal to three for adjacent cell edges as shown in Fig. 3.

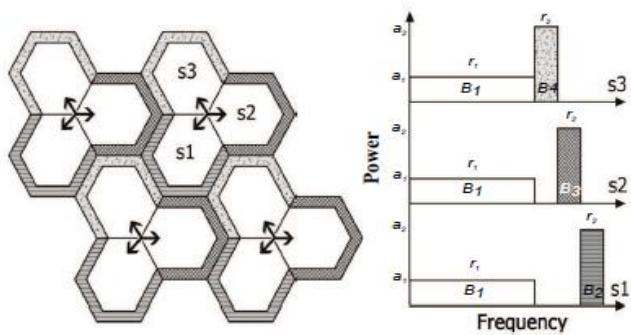

Figure. 3. Fractional Frequency Reuse (FFR)

## 2.1.2. SOFT FREQUENCY REUSE

The term soft reuse is due to the fact that effective reuse of the scheme can be adjusted by the division of powers between the frequencies used in the center and edge bands. SFR makes use of the concept of zone-based reuse factors in the cell-center and cell-edge areas. Unlike the PFR; however, frequency and power used in these zones are restricted. In particular, a frequency reuse factor of 1 is employed in the central region of a cell, while frequency reuse factor greater than 1 at the outer region of the cell close to the cell edge. In fact, when the mobile station is near the antenna of the base station, the received power of the wanted user signal is strong, and the interference from other cell is weak. So at the inner part of the cell, all the sub-carriers can be used to achieve high data rate communication. For example, consider the 3-sector cell sites shown in Fig. 4, the cell-edge band uses 1/3 of the available spectrum which is orthogonal to those in the neighboring cells and forms a structure of cluster size of 3. The cell-center band in any sector is composed of the frequencies used in the outer zone of neighboring sectors. [5]

The benefits of the soft frequency reuse scheme include the following [6]:
• Improved bit rate at cell edge;
• High bit rate at the cell center;
• Avoid interference at the cell edge, so the following procedure is easier: channel estimation, synchronization, cell selection and reselection.

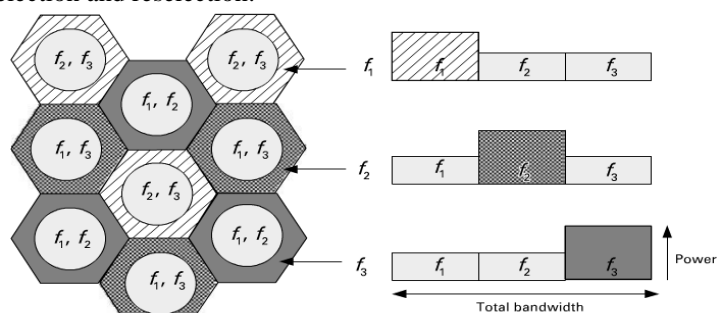

Figure. 4. Soft Frequency Reuse (SFR)

## 2.1.3. PARTIAL FREQUENCY REUSE

It is clear that using the same FRF value for the entire cell is not bandwidth-efficient [7]. One way to improve the cell-edge SINR, while maintaining a good spectral efficiency, is to use a frequency reuse factor greater than unity for the cell-edge regions and a reuse factor of unity for the cell-center regions [8]. In a homogeneous net-work, the cell center regions have equal areas. The idea of PFR is to restrict portion of the resources so that some frequencies are not used in some sectors at all. The effective reuse factor of this scheme depends on the fraction of unused frequency [9]. The PFR is also known as FFR with full isolation (FFR-FI), as users at cell-edge are fully protected (isolated) from adjacent cells interference [7]. An example for sites with 3 sectors is shown in Fig. 5.

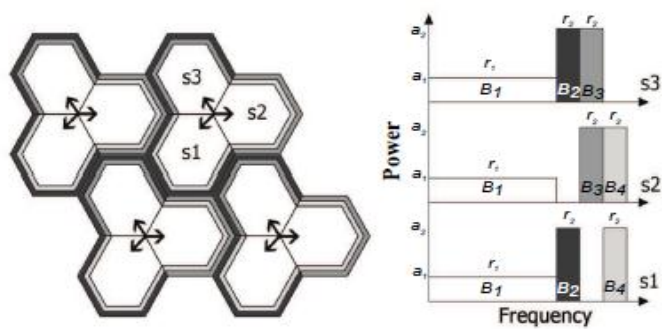

Figure. 5. Partial Frequency Reuse (PFR)

## 2.2. DYNAMIC SCHEMES
### 2.2.1. RELEASE 10: E-ICIC

In ICIC specifications, only homogeneous network scenario was examined. To deal with interference issue in heterogeneous deployments such as overlaying Macro and Femto cells, enhanced ICIC (e-ICIC) was standardized by 3GPP Release 10. Several enhancements were made to overcome the limitations of previous ICIC schemes. The e-ICIC mitigates interference on traffic and control channels. Besides, e-ICIC uses power, frequency and also time domain to mitigate intra-frequency interference in heterogeneous networks. In such networks, two major scenarios for severe inter cell interference should be highlighted: Macro-Pico scenario with Cell Range Extension (CRE) and Macro-Femto scenario with Closed Subscriber Group (CGS).

Extending the coverage of a cell by means of connecting a UE to cell that is weaker than the strongest detected cell is referred to as CRE. Indeed, cell selection in LTE is based on terminal measurements of the received power of the downlink signal. However; in a heterogeneous network we have different types of base stations that have different transmission powers including different powers of downlink signal. This approach for cell selection would be unfair to the low power nodes (Pico-eNBs) as most probably the terminal will choose the higher power base stations (Macro-eNBs) even if the path loss to the Pico-eNB is smaller and this will not be optimal in terms of uplink coverage, downlink capacity and interference. As a solution for the first 2 points cell selection could be dependent on estimates of the uplink path loss, which in practice can be done by applying a cell-specific offset to the received power measurements used in typical cell selection. This offset would somehow compensate for the transmitting power differences between the Macro-eNBs and Pico-eNBs; it would also ex-tend the coverage area of the Pico-eNB, or in other words extend the area where the Pico-eNB is selected. This area is called "Range Extension" and is illustrated in Fig. 6. [10]

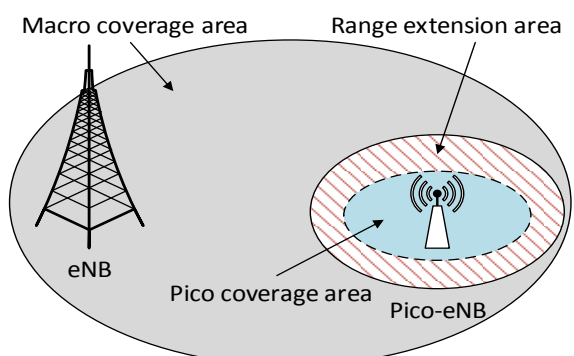

Figure. 6. Range extension illustration

Due to the difference in transmission powers of the Macro-eNBs and the Pico-eNBs, in the range extension area, illustrated in Figure 8, where the Pico-eNB is selected by the terminal while the downlink power received by that terminal from the Macro-eNB is much higher than the power it receives from the Pico-eNB, this makes the users in the range extension area more prone to interference from the Macro-eNB.

So along with the benefits of range extension comes the disadvantage of the high inter-cell interference that the Macro layer imposes on the users in the range extension area of the Pico layer. Fig. 7 illustrates the com-parison of 2 users connected to the Pico-eNB where:
- User 1 is placed close to the Pico-eNB so we will call it "center Pico user", this is not affected very much by the Macro-eNB interference as the downlink received power from the Pico-eNB is higher than the one received from the Macro-eNB.
- User 2 is placed farther from the Pico-eNB, in the range extension area, and as discussed before this user endures a severe interference from the Macro-eNB.

Solutions for the high interference levels in the range extension area will be discussed later. [11]

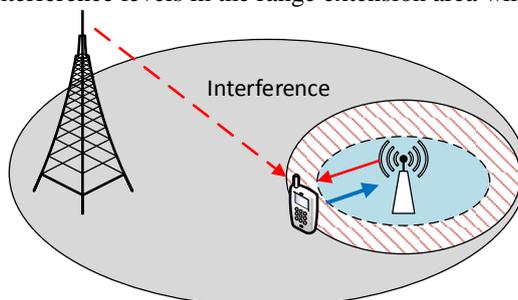

Figure. 7. Range extension interference

A closed subscriber group (CSG) is a limited set of users with connectivity access to a Femto cell. When a Femto cell is configured in CSG mode, only those users included in the Femto cell access control list are allowed to use the Femto cell resources.

When considering CSGs, yet another challenge for interference management arises if UEs are in the coverage area of a HeNB, typically well shielded from the macro-eNB, but are not allowed access to it. This creates complex high-interference scenarios in both transmission directions that cannot easily be solved. In the downlink such a "macro-UE" is the victim being exposed to heavy interference from the HeNB, whereas in the uplink the macro-UE is the aggressor severely disturbing transmissions to the HeNB (see Fig. 8).

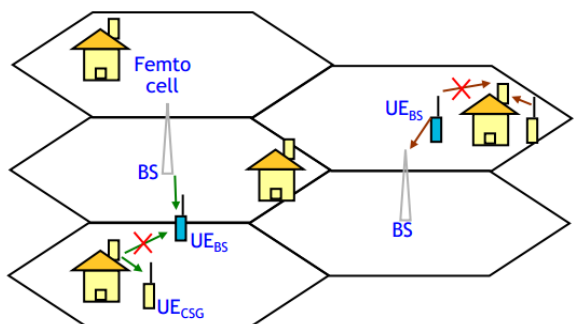

Figure. 8. Interference caused by CSG

Two schemes are newly introduced by Release 10: frequency domain e-ICIC based on Carrier Aggregation (CA) and time domain based e-ICIC called Almost Blank Subframes (ABS).
• CA based e-ICIC

The frequency domain e-ICIC manages radio resource, notably the radio resource blocks, such that multiple cells coordinate use of frequency domain resources. The main FDM interference cancellation method used in LTE-Advanced is carrier aggregation which is one of the most important features of LTE Advanced and it basically enables an LTE-Advanced user equipment (UE) to be connected to several carriers simultaneously. Carrier aggregation not only allows resource allocation across carriers but also allows scheduler based fast switching be-tween carriers without time consuming handovers, which means that a node can schedule its control information on a carrier and its data information on another carrier. An example of that concept in a HetNet scenario is to partition the available spectrum into, for example, 2 separate component carriers, and assign the primary component carrier (f1) and the second component carrier (f2) to different network layers at a time as shown in Fig. 9.

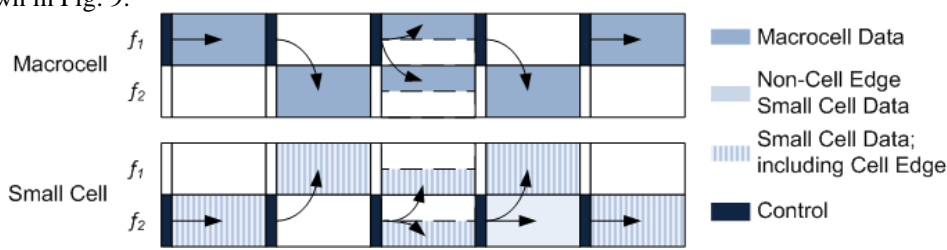

Figure. 9. Carrier Aggregation based e-ICIC

As shown Figure 10 the Macro layer can schedule its control information on f1 but can still schedule its users on both f1 and f2 so by scheduling control and data information for both Macro and Pico layers on different component carriers, interference on control and data can be avoided. It is also possible to schedule center Pico-eNB users12 data information on the same carrier that the Macro layer schedules its users as shown in the third subframe in Figure 10, as the interference from the Macro layer on center Pico-eNB users can be tolerated, while Pico-eNB users in the range extension areas are still scheduled in the other carrier where the Macro-eNB users are not scheduled.

The disadvantage of carrier aggregation with cross carrier scheduling is that it is only supported by release 10 terminals and onwards so this feature cannot be used by release 8 and 9 terminals.[11]
• ABS based e-ICIC

The e-ICIC introduces the concept of "Almost blank subframe" (ABS). The outline of ABS has been specified by the 3GPP in [12]. In this approach transmissions from Macro-eNBs inflicting high interference onto Pico-eNBs users are periodically muted (stopped) during entire subframes. This way, UEs connected to pico/femto cells can send their data during such ABS frames and avoid interference from macro cell (see Fig. 10). However this muting is not complete since certain control signals are still to be transmitted. These control channels have to be transmitted even in the muted subframes to avoid radio link failure or for reasons of back-wards compatibility.

The basic idea is to have some subframes during which the Macro-eNB is not allowed to transmit data allowing the range extension Pico-eNB users, who were suffering from interference from the Macro-eNB transmission, to transmit with better conditions. With time domain ICIC, a CRE UE may continue to be served by a victim cell (i.e., the weaker cell) even while under strong interference from aggressor cells (i.e., the stronger cell). A UE under strong interference from aggressor cells may need to mitigate interference from the aggressor cells on some physical channels and signals in order to receive data from serving cell or to detect the weak cells or to perform measurements on the weak cells.[13]

For the time domain ICIC, subframe utilization across different cells are coordinated in time through back-haul signalling or OAM configuration of so called ABS patterns. [13]

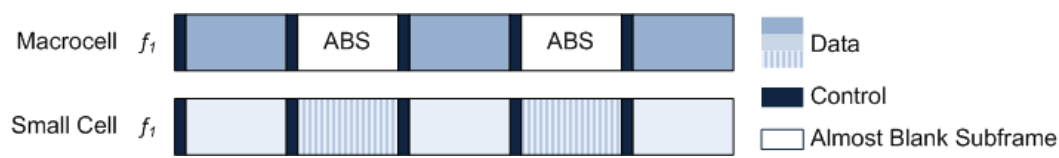

Figure. 10. Almost Blank Subframes based e-ICIC

Semi-static coordination techniques developed in LTE Rel-8/9/10 can provide benefits in managing inter cell interference with relatively little backhaul overhead and low implementation complexity. However, the achievable benefit is limited as it is not able to follow the dynamic interference condition caused by dynamic scheduling and beamforming. Moreover, semi-statically preserved resources result in low efficiency of resource utilization. Dynamically coordinating the transmission and reception of signals at multiple cells could lead to further performance improvement. Coordinated Multi Point transmission / reception (CoMP) was initially brought forward in the study item of LTE-Advanced to meet the requirements of IMT-Advanced in 2008. After further study, performance gain was identified, and a work item was approved to specify CoMP in LTE Rel-11. [5]

2.2.2. **RELEASE 11: COMP**

LTE-Advanced continues to evolve. New CA configurations are added (additions of new bands for CA are not bound to specific releases) and there are new features introduced in coming releases of the 3GPP specifica-tions, such as Coordinated Multi Point (CoMP) introduced in R11.

The main reason to introduce CoMP is to improve network performance at cell edges. The b   asic idea of CoMP is to transform inter cell interference into a useful signal, especially in cell edges where performance may be degraded.

In CoMP, a number of transmit points (TX) provide coordinated transmission in the downlink, and a number of receive points (RX) provide coordinated reception in the uplink. A TX/RX-point constitutes of a set of co-located TX/RX antennas providing coverage in the same sector. The set of TX/RX-points used in CoMP can ei-ther be at different locations, or co-sited but providing coverage in different sectors, they can also belong to the same or different eNBs. CoMP can be performed in several ways, and the coordination can be done for both homogenous networks as well as heterogeneous networks. When CoMP is used additional radio resources for signaling is required e.g. to provide UE scheduling information for the different DL/UL resources. [13]

There are two methods for CoMP technology deployment: the distributed control based on independent con-figuration of each eNB and the centralized control based on Radio Remote Heads (RRHs) also called Radio Re-mote Equipments (RREs).

The RRH concept constitutes a fundamental part of a state-of-the-art base station architecture. RRH-based system implementation is driven by the need to reduce both CAPEX and OPEX consistently, which allows a more optimized, energy-efficient, and greener base deployment. For example, the Open Base Station Architec-ture Initiative (OBSAI) and the Common Public Radio Interface (CPRI) standards introduced standardized interfaces separating the Base Station server and the remote radio head (RRH) part of a base station by an optical fiber.

With the distributed control, signaling is transmitted over X2 interface between eNBs to ensure inter cell co-ordination. Therefore, delay and overhead problems could be generated. However, in the centralized control, several RRHs are connected through optical fiber which transports base band signals between the different cells and the central eNB. Consequently, the radio resource management of all cells is exclusively performed by the central eNB which considerably reduces the eNBs configuration com-plexity. Nevertheless, the RRHs number should be limited in order to maintain a certain level of processing load at the central eNB.

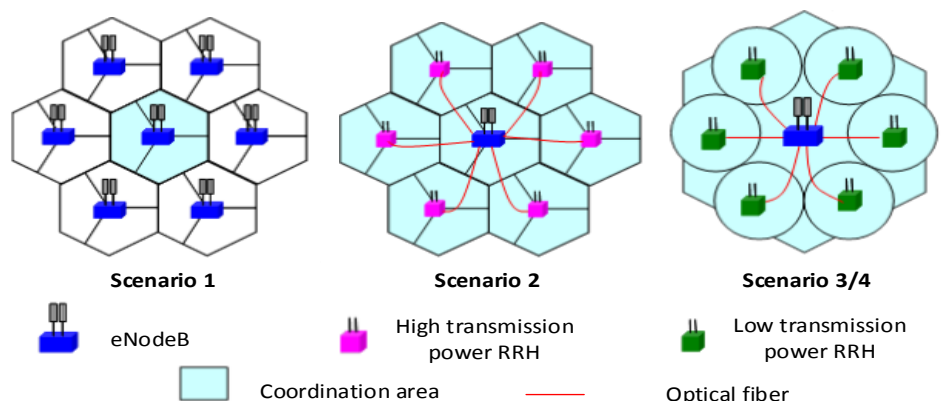

Figure. 11. CoMP scenarios selected by 3GPP

Four scenarios were selected in the 3GPP LTE Advanced CoMP study (see Fig. 12) to evaluate DL and UL COMP [14]:
- Homogeneous network with intra site CoMP
- Homogeneous network with high transmission power RRHs.
- Heterogeneous network with low transmission power RRHs within macro cell coverage. Each RRH forms a small cell with physical cell identity (PCI) independent from the macro cell.
- Heterogeneous network with low power RRHs within macro cell coverage where the transmission/reception points created by the RRHs have the same PCI as the macro cell.

The CoMP coordination area shown in Fig.12 is defined by the area containing the cooperating set. The co-operating set is a group of (geographically separated) points directly and/or indirectly participating in data transmission to a UE in DL or points that may be intended for data reception from a UE in UL in a time-frequency resource. Note that this set may or may not be transparent to the UE [15].

Different CoMP categories are used in DL and UL.

- Downlink COMP :

For DL CoMP, 3GPP standardization group has mainly focused on transmission schemes and potential im-pact on radio interface which gathers four essential areas: channel state information (CSI) feedback from UE, preprocessing schemes and reference signal design.
- Transmission schemes

Four categories are defined for DL CoMP transmission schemes in 3GPP framework: dynamic point selection (DPS), dynamic point blanking (DPB), joint transmission (JT), and coordinated scheduling/beamforming (CS/CB).

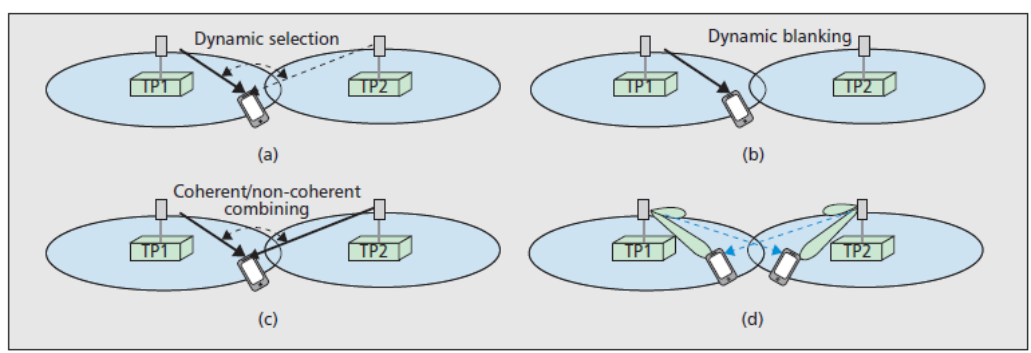

Figure. 12. Downlink CoMP transmission schemes

With DPS, data is available for transmission at two or more transmission points but only scheduled from one transmission point (TP) in each sub-frame. The TP is dynamically selected according to the instantaneous channel condition on per sub-frame basis. Indeed, The TP with the best link quality is selected to exploit channel variations opportunistically.

By DPB, the principal interferer(s) to UE in the coordination area is identified and dynamically muted (i.e., no signal is transmitted from this TP). By muting the dominant TP causing interference, SINR of the UE may be significantly improved as a few dominant interferers may represent a majority of the whole interference. A TP is muted only when there is gain on the utility of the whole coordination area. Muting the dominant interferer causes performance loss of the particular TP. If performance improvements of those beneficiaries outweigh the loss, the TP can be muted; otherwise, it should remain active to transmit d ta. It seems unfair to UE connected to the muted TP; however, since scheduling is dynamically performed, those UE devices may also benefit from muting other TPs in subsequent sub-frames. On the whole, it is beneficial to all UE if the scheduling algorithm is designed carefully [16].

JT is a form of spatial multiplexing that takes advantage of decorrelated transmission from more than one point within the cooperating set. Data to a UE is simultaneously transmitted from multiple points to coherently or non-coherently improve the received signal quality or data throughput. JT is categorized into coherent and non-coherent JT, depending on whether coherent combining of signals from multiple TPs is targeted. Coherent JT offers the best performance among all considered CoMP schemes, since it allows multiple TPs to completely nullify interference toward co-scheduled UE [16].

With CS/CB, data for a UE is only available at and transmitted from one point in the CoMP cooperating set but user scheduling and beamforming decisions are made across all points in the cooperating set. Scheduling a UE means to select a specific UE for transmission. The scheduling and beamforming among several TPs are aligned to decrease interference. Severe interference conditions may be alleviated with coordinated scheduling decisions. Furthermore, interference can be additionally reduced by an adequate choice of beamforming weights.

The ideal manner to optimize performance is to carry out scheduling and beam selection jointly. Nevertheless, this is not practical for the different coordination set. Consequently, an iterative procedure seems to be a simple and efficient solution.

- CSI feedback

Channel state information refers to known channel properties of a communication link. This information describes how a signal propagates from the transmitter to the receiver and represents the combined effect of, for example, scattering, fading, and power decay with distance. The CSI makes it possible to adapt transmissions to current channel conditions, which is crucial for achieving reliable communication with high data rates in multi antenna systems. CSI needs to be estimated at the receiver and usually quantized and fed back to the transmitter (although reverse-link estimation is possible in TDD systems). Therefore, the transmitter and receiver can have different CSI.

There are two main types of feedback: the explicit feedback which involves reporting the raw channel as observed by the receiver without assuming any processing, and the implicit feedback which involves reporting channel-related characteristics assuming hypotheses on a particular transmission/reception processing [17].

Different forms of explicit feedback are available in support of DL CoMP. They are all characterized by having a channel part and a noise-and-interference part. In implicit feedback, there are hypotheses at the UE and the feedback is based on one or a combination of two or more of them. One typical example of implicit feedback is the CQI/PMI/RI framework introduced in 3GPP Release8. In this case, CSI includes the precoding matrix indicator (PMI), rank indicator (RI), and channel quality indicator (CQI). The RI is the transmission rank or number of independent data streams that can be supported by the channel in spatial multiplexing transmission. The PMI is an index to a code word in a predefined codebook comprising precoding matrices. The CQI reflects the quality of the target link, which is a reference for modulation and coding scheme (MCS) selection.

A transmission hypothesis in CoMP is composed of two parts: signal hypothesis and interference hypothesis. The signal hypothesis specifies TP(s) from which the packet data is assumed to be transmitted, and the interference hypothesis stands for interference suffered during the assumed data transmission. CSI corresponding to one transmission hypothesis is defined as a CSI process. A CSI process is determined by the association of a signal hypothesis and an interference hypothesis, where the signal hypothesis and interference hypothesis are measured through CSI-RS and interference measurement resource (IMR), respectively. [16]

- Interference measurements

The UE should be able to measure the interference corresponding to the interference hypothesis in each CSI process. Release 11CoMP has special resource elements (RE) for interference measurement called interference measurement resource (IMR). Each CSI process is linked with a configured IMR. The IMR is defined by a number of occupied REs that are muted in certain TPs, so there is no transmitted signal from those TPs in the configured Res for IMR. The UE only receives signal on those REs to estimate interference from the other TPs.

- Reference signal design

CSI-Reference Signal or CSI-RS is used from the UE to calculate and report the CSI feedback (CQI/PMI/RI). Feedback of CSI is based on a separate set CSI-RS. CSI-RS are relatively sparse in frequency but regularly transmitted from all antennas at the base station. The CSI-RS is transmitted in each physical antenna port or virtualized antenna port and is used for measurement purposes only. A cell can be configured with one, two, four or eight CSI-RS. The exact CSI-RS structure, including the exact set of resource elements used for CSI-RS in a resource block, depends on the number of CSI-RS configured within the cell and may also be different for different cells. More specifically, within a resource-block pair there are 40 possible positions for the reference symbols of CSI-RS and, in a given cell, a subset of corresponding resource elements is used for CSI-RS transmission. A CSI process is associated with a CSI-RS to derive the signal hypothesis of the CSI process.

- Uplink COMP :

With UL multi cell reception, the signal from a UE is received by several cells and then combined. Contrarily to the DL, the UE has no need to know if a multi cell reception is performed. Consequently, UL CoMP should have little impact on the specifications of radio interface. Each UL CoMP scheme may be categorized into one of the following categories.

- Joint Reception (JR): signal transmitted by the UE is received jointly at multiple points (part of or entire CoMP cooperating set) at a time, e.g., to improve the received signal quality
- Coordinated Scheduling and Beamforming (CS/CB): user scheduling and precoding selection decisions are made with coordination among points corresponding to the CoMP cooperating set. Data is intended for one point only.

All CoMP schemes implementations introduce additional overhead. So there is clearly a trade-off between CoMP gain and backhaul overhead.

### III. COMPARATIVE STUDY

3GPP Release 8 LTE ICIC is a static method to decrease interference between neighboring macro base stations. This is done by lowering the power of a part of the sub-channels in the frequency domain which then can only be received close to the base station. These sub-channels do not interfere with the same sub-channels used in neighboring cells and thus, data can be sent faster on those sub-channels to mobile devices close to the cell.

3GPP Release 10 LTE-Advanced e-ICIC is part of the HetNet approach. While the macro cells emit long range high power signals, the small cells only emit a low power signal over short distances. To mitigate interference between a macro cell and several small cells in its coverage area, e-ICIC introduces frequency domain and time domain coordination schemes.

CoMP is however designed for both homogeneous and heterogeneous networks. This method lies on the coordination between different points to manage interference. The main idea behind this coordination scheme is the utilization of inter cell interference in a constructive manner.

In this part, we will present the simulation results to compare ICIC and e-ICIC methods at a first time. Later, we will refer to a recent study to compare e-ICIC with CoMP schemes.

As can be seen from Fig. 13, e-ICIC offers an enhancement of 80% for cell edge throughput compared to ICIC schemes. The simulation was performed for one Macro cell with two overlaid Pico cells. UEs are distributed uniformly with 25 UEs per Macro/ Pico cell.

It is also worth to notice that despite only Pico cells were used in the simulation, the same results are available for all other small cells types.

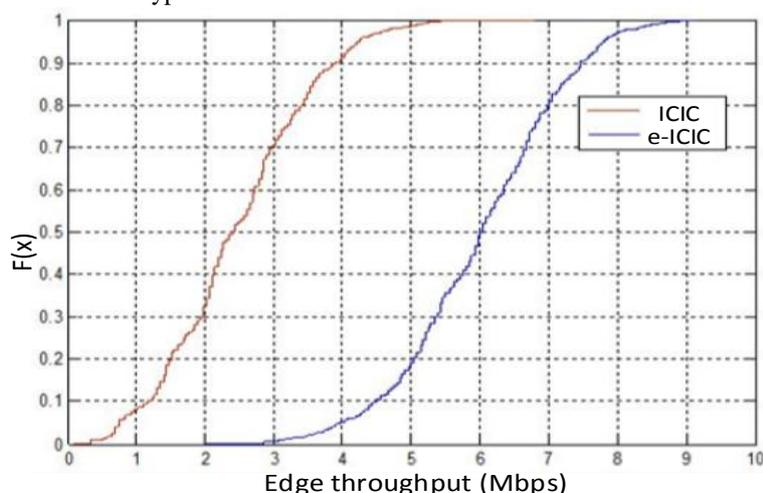

Figure. 13. Comparison between ICIC and e-ICIC schemes

The authors in [18] have studied the performance of e-ICIC scheme in comparison with CoMP and more specifically a combination of coherent JT and DPS schemes. The results presented in Fig. 8 show that CoMP scheme provides 25% throughput gain for cell edge users.

As mentioned before, coherent JT provides the best performance among all CoMP schemes. From the deep study carried in [18], we can conclude that a combination of two or more CoMP schemes is also possible and can lead to further ameliorations.

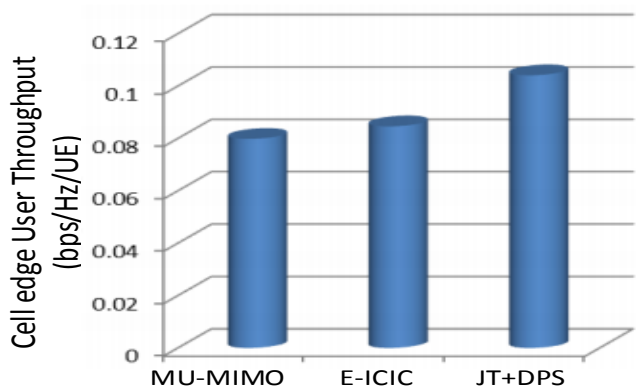

Figure. 14. Comparison between e-ICIC and CoMP schemes

## IV. CONCLUSION AND PERSPECTIVES

Deep research and standardization work was carried over LTE and LTE-Advanced networks. The evolution of interference mitigation mechanisms from Release 8 to Release 11 was presented in this paper. The different discussed schemes were classified into static and dynamic ones. Later, we provide a comparison between ICIC, e-ICIC and CoMP interference coordination methods. The obtained results have shown the performance enhancement in terms of cell edge throughput from Release 8 to Release 11 schemes.

For advanced networks beyond 4G, research and standardization work is not yet finished. 3GPP is working on new features in Release 12 and 13 to further resolve interference problems especially caused by emerging small cells.

In next works, we aim at developing a new self-organized interference mitigation method for Macro and small cells based on CoMP schemes and that will be completely compatible with 4G and beyond networks also known as 5G networks.

## REFERENCES

[1] http://www.3gpp.org/technologies/keywords-acronyms/98-lte
[2] 3GPP TS 36.300 V8.7.0 (2008-12)
[3] Huawei, "3GPP TSG RAN WG1 Meeting #41, R1-050507 – Soft Frequency Reuse Scheme for UTRAN LTE," 2005.
[4] Siemens, "3GPP TSG RAN WG1 Meeting #41, R1-050476: Evolved UTRA uplink scheduling and frequency reuse," 2005.
[5] A Survey on Inter-Cell Interference Coordination Techniques in OFDMA-based Cellular Networks, Abdelbaset S. Hamza, Shady S. Khalifa, Haitham S. Hamza, and Khaled Elsayed, Cairo University, March 19, 2013
[6] 3GPP TSG RAN WG1 Meeting #41, Athens, Greece, 9 – 13 May, 2005
[7] R. Kwan and C. Leung, \A Survey of Scheduling and Interference Mitigation in LTE," Electrical and Computer Engineering, vol. 2010, 2010.
[8] F. Khan, LTE for 4G Mobile Broadband: Air Interface Technologies and Performance.: Cambridge University Press, 2009.
[9] M. Rahman and H. Yanikomeroglu, \Enhancing cell-edge performance: a downlink dynamic interference avoid- ance scheme with inter-cell coordination," IEEE Transactions on Wireless Communications, vol. 9, no. 4, pp.1414-1425, 2010.
[10] RP-101229 RAN2 CRs on Core part: Enhanced ICIC for non-CA based deployments of heterogeneous networks for LTE RAN2; 3GPP,TSG RAN meeting #50
[11] Hisham El Shaer, Interference Management In LTE-Advanced Heterogeneous Networks Using Almost Blank Subframes, March 2012
[12] 3GPP, 3rd generation partnership project; Technical specification group radio access network; ''Way forward on time-domain extension of Rel 8/9 backhaul-based ICIC'', R1-105779, October 2010.
[13] 3GPP TS 36.300 V11.7.0, Group Radio Access Network; Evolved Universal Terrestrial Radio Access (E-UTRA) and Evolved Universal Terrestrial Radio Access Network (E-UTRAN); Overall description; Stage 2 (Release 11), September 2013
[14] 3rd Generation Partnership Project; Technical Specification Group Radio Access Network; Coordinated multi-point operation for LTE physical layer aspects (Release 11), September 2013, 3GPP TR 36.819 V11.2.0
[15] Christian F. Lanzani!, Georgios Kardaras†, Deepak Boppana ; Remote Radio Heads and the evolution towards 4G networks; February 2009
[16] Shaohui Sun, Qiubin Gao, Ying Peng, Yingmin Wang, Lingyang Song, "Interference Management Through Comp In 3gpp Lte-Advanced Networks", February 2013
[17] Afif Osseiran,Jose F. Monserrat,Werner Mohr, « Mobile and Wireless Communications for IMT-Advanced and Beyond", 2011
[18] Long Gao, Hitachi America, Santa Clara, Heterogeneous Networks – Theory and Standardization in LTE, IEEE WCNC 2013